\documentclass[prd,twocolumn,floatfix,showpacs]{revtex4}
\usepackage{hyperref}
\usepackage{epsfig}
\usepackage{color}
\usepackage{amsmath}
\newcommand{\mat}[1]{\mathbf{\mathrm{#1}}}
\begin{document}                
\hyphenation{wave-guide}
\pagestyle{myheadings} \markright{Elementary Particles as
Solutions of a 4-Dimensional Source Equation}
\title{Elementary Particles as Solutions of a 4-Dimensional Source Equation}
\author{Jos\'e B. Almeida}
\affiliation{Universidade do Minho, Departamento de F\'isica,
4710-057 Braga, Portugal.} \email{bda@fisica.uminho.pt}




\date{\today}
\begin{abstract}                
The author discusses particular solutions of a second order
equation designated by source equation. This equation is special
because the metric of the space where it is written is influenced
by the solution, rendering the equation recursive. The recursion
mechanism is established via a first order equation which bears
some resemblance to Dirac equation. In this paper the author
limits the discussion to solutions with constant norm but makes
use of 4-dimensional hypercomplex numbers in matrix
representation, a concept that is formally introduced in a section
devoted to that aspect. The particular solutions that are found
exhibit symmetries that can be assigned to spin, electric and
color charges of elementary particles, leaving mass as a free
parameter. Massless particles can also be assigned to special
solutions of the source equation, with the cases of photons,
gluons and gravitons clearly identified, together with another
massless particle which does not seem to be related to anything
detected experimentally. Another section deals with particle
dynamics under fields, showing that both gravitational and
electrodynamics can be modelled by geodesics of the spaces whose
metric tensors result from the recursion mechanism. Finally the
author suggests two lines of future work, one deriving fields from
densities and currents of masses and charges and the other one
aimed at determining particles' masses.

\end{abstract}

\maketitle
\normalcolor
\section{Background}
One can conceive of a very large computer with the sole task of
solving a small number of simultaneous differential equations,
with the only itch that the equations are \emph{evolutionary},
meaning that in the course of time their form evolves as a result
of previously found solutions. Assuming that the problem could be
appropriately formulated, this would create an evolutionary system
of equations whose complexity would surely outgrow the computing
power of any machine. In a science-fiction pass one could also
assume the computer to be itself a product of the equations, such
that its computing power would grow in parallel to the equations'
complexity. The author proposes that the Universe as a whole is
such a system, whose evolution can be understood up to our ability
to understand the basic equations. We will deal with evolutionary
equations by writing non-linear equations in 4-dimensional space;
the solutions to these will be seen as evolutionary when
3-dimensional space is considered. To this effect we will focus on
one particular form of the basic equations as such equations are
not known to us in general, although we can discuss some of their
characteristics. This particular form is designated by
\emph{source equation} in Ref.\ \cite{Almeida02:3} because it is
the source of dynamics and \emph{dynamic space}, an essential
concept introduced in the reference cited earlier. We believe the
source equation to be a condensation of the basic equations in a
similar way to which Klein-Gordon and Helmholtz equations are
condensations of Dirac and Maxwell equations \cite{Cottingham98,
Born80}. The reader may also find useful a look at Ref.\
\cite{Almeida02:1} as this is a precursor of the present paper,
establishing most of the assumptions used here, including setting
the problem in a very specific 4-dimensional space, with the
justification for such an approach.

This discussion is about a mathematical problem, i.\ e.\ without
any preexisting physical assumptions: All the variables in the
equations are pure numbers, unless clearly stated otherwise, so
the question of dimensional homogeneity is never an issue. This
facet of the issues at hand is detailed in Ref.\
\cite{Almeida02:3} where it was shown that dimensions and physical
interpretation can be introduced at a later stage, simultaneously
with the definition of the physical constants $c$, $G$, $\hbar$
and $e$.

The general form of the source equation is
\begin{equation}
    \label{eq:founding}
    g^{\mu \nu}
    \nabla_{\mu \nu} \psi = -\psi,
\end{equation}
where $g^{\mu \nu}$ is a space metric, $\psi$ is a function whose
mathematical characteristics will be discussed in the following
section, and where $\nabla_\mu$ designates the covariant
derivative with respect to the coordinate $x^\mu$. We use
Einstein's summation convention, with Greek indices and
superscripts taking values between 0 and 3 and Roman ones running
from 1 to 3. It is presumed that the basic equations of the
dynamics will provide a means of determining $g^{\mu \nu}$ from
$\psi$, in such a way that the successive equation-solution sets
resulting from Eq. (\ref{eq:founding}) become coupled in a
recursive manner, i.\ e.\ the solutions become self sustained.
This process then links the space metric to the dynamics, defining
what we have designated as \emph{dynamic space}, a space which
contains and is generated by the dynamics. The space metric will
be considered separable in two components: an ``inertia"
reflecting the influence of the particular point on the dynamic
space whose coordinates are under consideration, and a ``field"
reflecting the influence of all other points.

It is important to emphasize that dynamic space excludes anything
that does not participate in the dynamics under study. If we take
the viewpoint of Physics for one moment, at the Universe's scale
everything participates in the grand dynamics and is contained in
the Universe's dynamic space. When we decide to study some
particular dynamics isolated from the rest of the Universe, this
dynamics defines a space for itself, which excludes everything
else. Influences from other dynamic spaces can be accounted for
via the field component of the metric. Consider the Earth's orbit
around the Sun: We have the choice of using the dynamic space
defined by the two bodies, or the dynamic space of the Earth
alone, influenced then by a field component due to the Sun. These
two approaches are equivalent if within the dynamic space concept
we take universal space as a superposition of dynamic spaces,
within which we can choose the components for the need of a
particular study.

In connection with dynamic space within the above view of
superposed spaces, we define also an \emph{observer space}, which
is by definition obtained through a coordinate change that removes
the inertia component from the metric of the dynamic space under
consideration by the observer. Observer space can be then seen as
the result of a continuum of superposed dynamic spaces where each
point can be mapped to a point of one particular dynamic space via
inserting an inertia component in the dynamics equation of that
space (as discussed in Ref.\ \cite{Almeida02:3}). In Physics,
imagine one galaxy as dynamic space: When the galaxy's dynamics is
studied in isolation its dynamic space extends to infinity. With
the inertia component related to mass density, and with this
density tending to zero with distance, the conversion to observer
space confines the galaxy to a limited region of observer space. A
similar link between dynamic and observer spaces can be applied to
planetary systems, to the atom, and eventually to particles.

This paper is the first in a series which will examine some
mechanisms of dynamic space definition in a gradual process which
has the potential to lead to universal space, starting here with
``elementary" particles. Next section deals with the mathematical
characteristics of the function $\psi$ as the solution of
Helmholtz equation, also known as the ``standing wave" equation,
itself identified as contained in the source equation. We examine
first 2- and 3- dimensional settings, with the ultimate intent to
obtain a 4-dimensional formalism adequate for expressing the
source equation solutions and metric. Once such solutions are
found, their physical interpretation with and without fields
follows to confirm their adequacy by leading to key features of
gravitation and electromagnetism. Particles involved with
chromodynamics will only be introduced, without the necessary
dynamic details.
\section{4-dimensional complex numbers}
A complex number can be represented in the exponential form by
\begin{equation}
    \label{eq:expcomplex}
    \psi = m \mathrm{e}^{\mathrm{i} \theta}
\end{equation}
or equivalently in real-imaginary form as
\begin{equation}
    \label{eq:complex}
    \psi = m(\cos \theta + \mathrm{i} \sin \theta),
\end{equation}
with $m$ a positive number called the norm or modulus of the
complex number. Either form is well suited to represent harmonic
functions of one variable relative to another by taking just the
real part. In the second form the variable $\psi$ as a harmonic
function would be dependent on the variable $\theta$ by the
relation $\psi = m \cos \theta$, with the first form representing
solutions
\begin{equation}
    \label{eq:complexsol}
    \psi = m \mathrm{e}^{\pm\mathrm{i} vx}
\end{equation}
of Helmholtz equation in one dimension
\begin{equation}
    \label{eq:1DHelmholz}
    \frac{\mathrm{d}^2}{\mathrm{d}x^2}\,\psi = -v^2\psi,
\end{equation}
with $\theta = vx$.

A complex number is also appropriate for representing
2-dimensional rotations since such rotations are harmonic
relations between two orthogonal directions of space. In this
interpretation a variation of coordinates in one of the directions
is represented by the real part, while the coefficient of the
imaginary part deals with the orthogonal direction. If the two
coordinates are associated as a complex number
\begin{equation}
    \label{eq:complexforms}
   x = x^1+ \mathrm{i}x^2,
\end{equation}
the rotated coordinates are given by
\begin{equation}
    \label{eq:2drot1}
    x' = \psi x,
\end{equation}
which can be expressed in complex form as,
\begin{equation}
    \label{eq:2drot2}
   x' = x^1\,\cos \theta - x^2\,\sin\theta + \mathrm{i}
   \left(x^1\,\sin \theta + x^2\,\cos \theta\right).
\end{equation}
Complex numbers with norm unity belong to a group usually
designated by $U(1)$, isomorphic to the group of 2-dimensional
rotations $SO(2)$, whereby rotations in opposite directions are
represented by plus and minus signs added to the exponent of their
exponential form as shown in Eq.\ (\ref{eq:expcomplex}), and a
$2\pi$ rotation in either direction is equivalent to no rotation
at all, i.\ e.\ the null rotation, with the exponential replaced
by unity, and is commonly referred as a ``full'' rotation.

In order to generalize complex numbers we designate by
\emph{imaginary unit} any \emph{traceless} matrix
$\mathbf{\mathrm{u}}$ that verifies the condition
\begin{equation}
    \label{eq:imunit}
    \mathbf{\mathrm{u}}^2 = -\mathbf{\mathrm{I}},
\end{equation}
then an \emph{imaginary matrix} is defined as the product of a
real number by an imaginary unit; the following relation is always
verified \cite{Dixon94}:
\begin{equation}
\begin{split}
    \label{eq:unithyper}
    \mathrm{e}^{\theta \mathbf{\mathrm{u}}} &=
    \frac{\mathrm{e}^{\theta \mathbf{\mathrm{u}}}
    +  \mathrm{e}^{-\theta \mathbf{\mathrm{u}}}}{2} +
    \frac{\mathrm{e}^{\theta \mathbf{\mathrm{u}}}
    -  \mathrm{e}^{-\theta \mathbf{\mathrm{u}}}}{2} \\
    &= \mathbf{\mathrm{I}}\,\cos \theta
    + \mathbf{\mathrm{u}}\,\sin \theta.
\end{split}
\end{equation}
In a similar formal manner, we can define also traceless matrices
$\mat{h}$ designated \emph{Hermitian units} such that $\mat{h}^2 =
\mathbf{\mathrm{I}}$. For such matrices we have
\begin{equation}
\begin{split}
    \label{eq:hermitunit}
    \mathrm{e}^{\theta \mat{h}} &=
    \frac{\mathrm{e}^{\theta \mat{h}}
    +  \mathrm{e}^{-\theta \mat{h}}}{2} +
    \frac{\mathrm{e}^{\theta \mat{h}}
    -  \mathrm{e}^{-\theta \mat{h}}}{2} \\
    &= \mathbf{\mathrm{I}}\,\cosh \theta
    + \mat{h}\,\sinh \theta.
\end{split}
\end{equation}

Rotations in 3 dimensions are associated with hypercomplex numbers
called quaternions (or 3D complex numbers) seen as an extension of
complex numbers to 3-dimensional space \cite{Altmann86}. We will
define them through the Pauli matrices:
\begin{equation}
    \label{eq:pauli2}
    \sigma^1 = \begin{pmatrix}
    0 & 1 \\ 1 & 0 \end{pmatrix},~~
    \sigma^2 = \begin{pmatrix}
    0 & -\mathrm{i} \\
    \mathrm{i} & 0 \end{pmatrix},~~
    \sigma^3 = \begin{pmatrix}
    1 & 0 \\
    0 & -1 \end{pmatrix},
\end{equation}
complemented with $\mat{I}$ as the identity matrix. They verify
the well-known unitary and anti-commutator relations
\begin{equation}
    \label{eq:paulirel}
    \sigma^i \sigma^j + \sigma^j \sigma^i = 2 \delta^{ij}\mathbf{\mathrm{I}},
\end{equation}
with $\delta^{ij}$ being the Kronecker delta, making them
Hermitian units as well as imaginary units when multiplied by
$\mathrm{i}$.

When multiplied by $\mathrm{i}$ they do form a quaternion
representational basis as defined by Hamilton \cite{Waerden91} if
the quaternion product is simultaneously defined by:
\begin{equation}
    \label{eq:quatprod}
    \mathrm{i}\sigma^j \otimes \mathrm{i}\sigma^k =
    \begin{cases}
    \sigma^j \sigma^k, & j \neq k, \\
    -\mat{I}, & j = k.
    \end{cases}
\end{equation}

A unitary $2 \times 2$ matrix representing unitary 3D complex
numbers  has the form
\begin{equation}
    \label{eq:quaternion}
    \psi = \mathrm{e}^{\mathrm{i}a_j \sigma^j},
\end{equation}
with the $a_j$'s real number components of a vector
$\bar{\mathbf{\mathrm{a}}}$ with length $\theta = (\delta^{ij}a_i
a_j)^{1/2}$. From Eq.\ (\ref{eq:unithyper}), Eq.\
(\ref{eq:quaternion}) is equivalent to
\begin{equation}
    \label{eq:altquaternion}
    \psi =\mathbf{\mathrm{I}}\,\cos \theta +
    \mathrm{i}\frac{a_j \sigma^j}{\theta}\,\sin \theta ,
\end{equation}
which is a unitary quaternion, with the identity matrix $\mat{I}$
as the 0th order of its matrix representation. It is associated
with a rotation of an angle $2\theta$ about the direction of unit
vector $\bar{\mathbf{\mathrm{a}}}/\theta$, as we shall see below.

Defining the Hermitian coordinate matrix of a point $(x^1,\,
x^2,\, x^3)$ in 3D via the Pauli matricial basis
\begin{equation}
    \label{eq:coordmatrix}
    \mathbf{\mathrm{x}} = \mathrm{i}\sum  x^j \sigma^j = \left(\begin{array}{cc}
      \mathrm{i}x^3 & x^2 + \mathrm{i}x^1 \\
      -x^2 + \mathrm{i}x^1 & -\mathrm{i}x^3 \
    \end{array}\right),
\end{equation}
we can see that the coordinates remain on a sphere of squared
radius $\mathrm{det}\,\mathbf{\mathrm{x}}= \delta_{ij} x^i x^j$
when transformed by the matricial similarity transformation
\begin{equation}
    \label{eq:rotation}
    \mat{x'} = \psi \mat{x} \psi^{-1} = \psi \mat{x} \psi^\dagger,
\end{equation}
where $\psi^\dagger$ is the Hermitian conjugate of $\psi$; notice
that the second equality results from $\psi$ being a unitary
matrix. The transformation above performs a generic rotation in 3D
containing the earlier 2D rotations with \emph{double their angle}
as we can see by taking a vector $\bar a$ along $\sigma^1$ (only
$a_1$ is non-zero) and using Eqs.\ (\ref{eq:altquaternion}) and
(\ref{eq:coordmatrix}) in the similarity Eq.\ (\ref{eq:rotation}).
Coordinate $x^1$ is then unchanged, and the other coordinates
transform as follows:
\begin{equation}
\begin{split}
    \label{eq:doublerot}
    {x'}^2 &= x^2 \cos(2 \theta) - x^3 \sin(2 \theta),\\
    {x'}^3 &= x^2 \sin(2 \theta) + x^3 \cos(2 \theta).
\end{split}
\end{equation}
The same relation holds true when a circular permutation of the
indices is performed. This result means that a $4\pi$ rotation
angle is induced or two full rotations, for one cycle of the
exponent \cite{Cottingham98, Sternberg95, Wigner59}. The group of
$2 \times 2$ unitary matrices defined by Eq.\
(\ref{eq:quaternion}) is designated by $U(2)$, with $SU(2)$ being
the sub-group of those matrices whose determinant is unity
\cite{Sternberg95, Cottingham98} and the group of real orthogonal
matrices of dimension 3, representing 3D rotations, is designated
by $SO(3)$. It is said that $SU(2)$ is a \emph{double cover} or
\emph{two-fold covering} of $SO(3)$ in view of the angle doubling
referred above \cite{Lounesto01}.

Generalization of complex numbers to 3 dimensions implies that we
consider also non-unitary matrices; here the situation is more
complicated than in 2 dimensions because the entities we are
defining are $2 \times 2$ matrices and so must also be whatever
generalizes the modulus of a standard complex. Eq.\
(\ref{eq:quaternion}) can be generalized with the insertion of a
traceless Hermitian matrix before the exponential; this factor is
designated as modulus of the 3D complex, which we will represent
by $|\psi|$:
\begin{equation}
    \label{eq:gen3dcomp}
    \psi = |\psi| \mathrm{e}^{\mathrm{i}a_j \sigma^j}.
\end{equation}
The modulus relates to the norm $||\psi||$ through
\begin{equation}
    \label{eq:modulus}
    |\psi|^2 = \psi^\dagger \psi = |\psi|^\dagger |\psi| = ||\psi||^2
\end{equation}
We will impose a further restriction on the modulus by taking the
norm as a positive real number:
\begin{equation}
    \label{eq:modrelation}
    |\psi|^\dagger |\psi| = m^2,
\end{equation}
Now we are ready to verify that the 3D complexes defined through
Eq.\ (\ref{eq:gen3dcomp}) can be used as solutions of a
3-dimensional Helmholtz equation. Consider the following equation
with $\psi$ taken as a standard complex number function of
coordinates $x^j$ and $\bar{\mathbf{\mathrm{v}}}$ a vector of
components $v_j$ and norm $v$:
\begin{equation}
    \label{eq:3dwave}
    \delta^{j k} \partial_{j k} \psi = - v^2\psi;
\end{equation}
usually the solution to this equation is written
\begin{equation}
    \label{eq:3dsol1}
    \psi = |\psi| \mathrm{e}^{\mathrm{\pm i}v_j x^j},
\end{equation}
where we have ignored an arbitrary phase factor and
\begin{equation}
    \label{eq:v2}
    v^2 = \delta^{jk}v_j v_k.
\end{equation}
This solution represents a plane \emph{standing wave} with
$\bar{\mathbf{\mathrm{v}}}$ seen as defining the orientation of
the wave. We call \emph{standing wave} to a standing harmonic
3-dimensional pattern. Imagine that $\psi$ represents light
intensity: The space will then be seen as alternating between
bright and dark along the wave direction; by defining planes
normal to that direction spaced at half the wavelength we could
split 3-dimensional space into alternating bright and dark zones.
Using instead $\psi$ as a 3D complex we can write different
solutions for the same equation (\ref{eq:3dwave}):
\begin{equation}
    \label{eq:3dsol2}
    \psi = |\psi| \mathrm{e}^{\pm \mathrm{i}v_j \sigma^j x^j}.
\end{equation}

We must verify that Eq.\ (\ref{eq:3dsol2}) together with Relations
(\ref{eq:modrelation}) and (\ref{eq:modulus}) is a solution of Eq.
(\ref{eq:3dwave}). The partial derivatives of $\psi$ are of the
form
\begin{equation}
    \label{eq:partderiv}
    \partial_j \psi =  \pm |\psi|\mathrm{i} v_j  \sigma_j
    \exp(\pm \mathrm{i}v_k \sigma^k x^k) =
    \pm \mathrm{i}v_j \sigma_j \psi
\end{equation}
where we have lowered the index of $\sigma_j$ in order to avoid
summation over $j$. The second derivative in $x^j$ of $\psi$
produces a term $\sigma_j \sigma_j$, and considering Relations
(\ref{eq:paulirel}), we get
\begin{equation}
    \label{eq:secderiv}
    \partial_{jj} \psi = - \left(v_j\right)^2 \psi.
\end{equation}
When the three second derivatives are added we obtain Condition
(\ref{eq:v2}), and thus Eq.\ (\ref{eq:3dwave}) is verified.

With either complex or 3D complex solutions for $\psi$ the general
solution of Eq.\ (\ref{eq:3dwave}) is obtained through a
superposition of positive and negative exponent solutions with
different phase factors. The types of solutions, though, are
different: Both equations represent standing plane waves but Eq.\
(\ref{eq:3dsol2}) is a associated to spinning of the points along
the wave direction at twice the wave frequency; we will now call
this a \emph{spin 1/2 wave}. General solutions with 3D complex
numbers are obtained by superposition of spin 1/2 waves; the norm
of the 3D complex solution is interpreted as the wave amplitude,
while the modulus contains information about its polarization.
Spin 1/2 waves are especially indicated to describe axially guided
waves, while no-spin waves, described by complex numbers, are
better suited to describe unguided waves.

The question remains about the possibility of defining
hypercomplex numbers representing solutions of Helmholtz equation
in 4 dimensions. It is impossible to add a 4th imaginary unit on a
par with the 3 quaternion units we used and end up with a division
algebra \cite{Dixon94}; this does not mean, however, that we
cannot define 4-dimensional complex numbers. We will start with a
discussion about 4-dimensional rotations, since in 2 and 3
dimensions complex numbers have been associated to rotations and
we are led to expect a similar relationship in 4D.

Rotations in n-dimensional space are isomorphic to group $SO(n)$
and determined by $n(n-1)/2$ parameters \cite{Sternberg95,
Wigner59, Lounesto01}, usually chosen as rotation angles; this
means that in 4 dimensions we have to select 6 different angles to
characterize all possible rotations. There are 6 very special 4D
rotations which preserve two of the coordinates while rotating the
other two. For instance the orthogonal matrices inducing rotations
that preserve $x^0, x^3$ or $x^1,x^2$ are respectively
\begin{gather}
    \label{eq:rot1}
    \left(\begin{array}{cccc}
      1 & 0 & 0 & 0 \\
      0 & \cos \theta & -\sin \theta & 0 \\
      0 & \sin \theta & \cos \theta & 0 \\
      0 & 0 & 0 & 1 \
    \end{array}\right),\\
    \label{eq:rot2}
    \left(\begin{array}{cccc}
      \cos \theta & 0 & 0 & -\sin \theta \\
      0 & 1 & 0 & 0 \\
      0 & 0 & 1 & 0 \\
      \sin \theta & 0 & 0 & \cos \theta \
    \end{array}\right).
\end{gather}
When right multiplied by the column vector of the coordinates
these special matrices rotate two coordinates by an angle $\theta$
while preserving the other two coordinates.

Group $SO(4)$ is doubly covered by group $SU(2) \times SU(2)$ in a
similar way to the double covering of $SO(3)$ by $SU(2)$
\cite{Lounesto01}; this suggests that we can replace Pauli
$\sigma$ matrices by Dirac matrices built from them through matrix
direct products. There are 15 of such matrices excluding the
identity which can be generated by the formulas
\begin{equation}
\begin{split}
    \label{eq:diracgen}
    \lambda^j &= \mat{I}_2 \odot \sigma^j,\\
    \rho^j &= \sigma^j \odot \mat{I}_2,
\end{split}
\end{equation}
where $\mat{I}_2$ is the $2 \times 2$ identity matrix and $\mat{A}
\odot \mat{B}$ is the matrix \emph{direct product}. Explicitly
this set of Dirac matrices is given by
\begin{alignat*}{2}
    \label{eq:diracsigma}
    \lambda^1 &= \left(\begin{array}{cccc}
      0 & 1 & 0 & 0 \\
      1 & 0 & 0 & 0 \\
      0 & 0 & 0 & 1 \\
      0 & 0 & 1 & 0 \
    \end{array}\right),&~~
    \lambda^2 &= \left(\begin{array}{cccc}
      0 & -\mathrm{i} & 0 & 0 \\
      \mathrm{i} & 0 & 0 & 0 \\
      0 & 0 & 0 & -\mathrm{i} \\
      0 & 0 & 0 & \mathrm{i} \
    \end{array}\right),\\
    \lambda^3 &= \left(\begin{array}{cccc}
      1 & 0 & 0 & 0 \\
      0 & -1 & 0 & 0 \\
      0 & 0 & 1 & 0 \\
      0 & 0 & 0 & -1 \
    \end{array}\right),&~~
    \rho^1 &= \left(\begin{array}{cccc}
      0 & 0 & 1 & 0 \\
      0 & 0 & 0 & 1 \\
      1 & 0 & 0 & 0 \\
      0 & 1 & 0 & 0 \
    \end{array}\right),\\
    \rho^2 &= \left(\begin{array}{cccc}
      0 & 0 & -\mathrm{i} & 0 \\
      0 & 0 & 0 & -\mathrm{i} \\
      \mathrm{i} & 0 & 0 & 0 \\
      0 & \mathrm{i} & 0 & 0 \
    \end{array}\right),&~~
    \rho^3 &= \left(\begin{array}{cccc}
      1 & 0 & 0 & 0 \\
      0 & 1 & 0 & 0 \\
      0 & 0 & -1 & 0 \\
      0 & 0 & 0 & -1 \
    \end{array}\right).
\end{alignat*}
The complete set of 16 Dirac matrices is generated by
\begin{equation}
    \label{eq:16dirac}
    \mat{E}^{\mu \nu} = \rho^\mu \lambda^\nu,
\end{equation}
where $\lambda^0 = \rho^0 = \mat{I}$.

The set of 16 Dirac matrices describes 15-dimensional space in the
same way as the set of 4 Pauli matrices describes 3-dimensional
space. The source equation and the 4D Helmholtz equation don't
require this high dimensionality because they can be associated to
4D rotations and hence to group $SU(2) \times SU(2)$, which can be
represented by Dirac matrices if appropriate relations are
introduced between its elements, effectively reducing the
dimensionality from 15 to 4. The problem we will be facing is more
complex, however because the source equation is obtainable from a
first order equation in much the same way as Klein-Gordon equation
can be obtained from Dirac equation; the process is the same that
was used in Ref.\ \cite{Almeida02:3}. From the cited reference we
recover the first order equation, here in a slightly modified form
\begin{equation}
    \label{eq:1storder}
    \mathrm{i}\mathbf{\mathrm{s}}^\mu
    \nabla_\mu  \psi
    = \psi.
\end{equation}
where $\mathbf{\mathrm{s}}^\mu$ are 4 $\times$ 4 Hermitian
matrices defined by
\begin{equation}
    \label{eq:recursion}
    \mathbf{\mathrm{s}}^0 = \frac{|\psi|}{||\psi||^2},~~~~
    \mathbf{\mathrm{s}}^j = \frac{\alpha^j}{||\psi||};
\end{equation}
where we define 4 $\alpha^\mu$ matrices by
\begin{equation}
    \label{eq:alpha}
    \alpha^0 = \mat{E}^{10},~~~~\alpha^j = \mat{E}^{3j},
\end{equation}
which verify the relation
\begin{equation}
    \label{eq:alpharel}
    \alpha^\mu \alpha^\nu + \alpha^\nu \alpha^\mu = 2 \delta^{\mu
    \nu}.
\end{equation}
Relation (\ref{eq:modulus}) is used for the construction of
$\mat{s}^\mu$. This is a more general procedure than was used in
Ref.\ \cite{Almeida02:3}.

Applying the operator $\mathrm{i}\mathbf{\mathrm{s}}^\mu
\nabla_\mu$ to both members, and assuming the condition
\begin{equation}
    \label{eq:genrieman}
    \nabla_\sigma \mathbf{\mathrm{s}}^\mu=0,
\end{equation}
the space metric of Eq. (\ref{eq:founding}) generated by the
$\mathbf{\mathrm{s}}^\mu$ matrices is:
\begin{equation}
    \label{eq:smetric}
    \mathbf{\mathrm{g}}^{\mu \nu} = \mathbf{\mathrm{s}}^\mu
    \mathbf{\mathrm{s}}^\nu.
\end{equation}
The condition imposed by Eq.\ (\ref{eq:genrieman}) automatically
verifies Riemann's condition $\nabla_\sigma g^{\mu \nu}=0$
although the converse is not necessarily true.

There is an apparent inconsistency in Eq.\ (\ref{eq:smetric})
because $\mathbf{\mathrm{s}}^\mu \mathbf{\mathrm{s}}^\nu \neq
\mathbf{\mathrm{s}}^\nu \mathbf{\mathrm{s}}^\mu$ and thus the
metric is defined as a non-symmetric tensor of $4 \times 4$ matrix
elements. However, in many circumstances, as will be the case in
this section, the metric will be anti-symmetric and cancellation
between corresponding terms will allow a diagonal metric to be
used.

Consider here expressions of the type $g^{\mu \nu} \nabla_{\mu
\nu}$, which imply an addition over all the $\mu \nu$
combinations; since the terms $\mu \nu$ and $\nu \mu$ are
symmetric and cancel each other, the expression is equivalent to
what would be obtained with a diagonal metric. In index raising or
lowering operations it is not indifferent whether the metric is
symmetric or anti-symmetric; we will always consider the metric to
be defined by Eq.\ (\ref{eq:smetric}) followed by elimination of
anti-symmetric elements, making this elimination an integral part
of dynamic space construction.

In order to verify Eq.\ (\ref{eq:1storder}) and considering
relations (\ref{eq:recursion}) we must try solutions of the type
\begin{equation}
    \label{eq:4dquaternion}
    \psi = \psi_0 \mathrm{e}^{\mathrm{i}p_j \alpha^j x^j},
\end{equation}
where we have used $\psi_0$ to denote the modulus and $x^0$
dependence of $\psi$. Solutions of the form given above  can be
inserted into Eq.\ (\ref{eq:1storder}) resulting in
\begin{equation}
    \label{eq:3terms}
    \frac{\mathrm{i}|\psi|\nabla_0 \psi}{||\psi||^2}
    -\frac{\sum a_j \psi}{||\psi||} = \psi,
\end{equation}
leaving only the $x^0$ dependence to be contended with.

We will define 4D complex numbers to be $4 \times 4$ matrices,
solutions of Eq.\ (\ref{eq:1storder}), of the form
\begin{equation}
    \label{eq:4dcomplex}
    \psi = |\psi| \mathrm{e}^{p_\mu \mat{u}^\mu x^\mu},
\end{equation}
where $\mat{u}^j = \mathrm{i}\alpha^j$ and $\mat{u}^0$ is some
imaginary unit independent of the other 3. If we were only
concerned with the simplified source equation above any Hermitian
matrix with a trace of $4m$ could be used as modulus; this does
not occur when we consider also Eq.\ (\ref{eq:1storder}). For
simplicity we start by examining the situations where $p_0 = m$,
$p_j = 0$; equation (\ref{eq:1storder}) is simplified to
\begin{equation}
    \label{eq:1stordstat}
    \mathrm{i}\mathbf{\mathrm{s}}^0 \nabla_0 \psi = \psi.
\end{equation}
Considering the definition of $\mathbf{\mathrm{s}}^0$ in Eq.\
(\ref{eq:recursion}) and noting that the covariant derivative of
$\psi$ is equal to its partial derivative
\begin{equation}
    \label{eq:1stordstat2}
    \mathrm{i}|\psi| m \mathbf{\mathrm{u}}^0 \psi = m^2 \psi,
\end{equation}
where we have made $||\psi|| = m$ and multiplied both members by
$m^2$ to remove the denominator. This equation is verified if
\begin{equation}
    \label{eq:cond0}
    |\psi| =  \mathrm{i} m \mathbf{\mathrm{u}}^0.
\end{equation}

We are now in position to remove the simplifying condition
inserted above and write the general solution
\begin{equation}
    \label{eq:nonstatpart}
    \psi = \mathrm{i} m \mathbf{\mathrm{u}}^0 \mathrm{e}^{s
    p_\mu \mathbf{\mathrm{u}}^\mu x^\mu},
\end{equation}
where $s$ is a sign factor with the values $\pm1$ and necessarily
Condition (\ref{eq:4dcond}) is verified.

\begin{figure}[htb]
    \centerline{\psfig{file=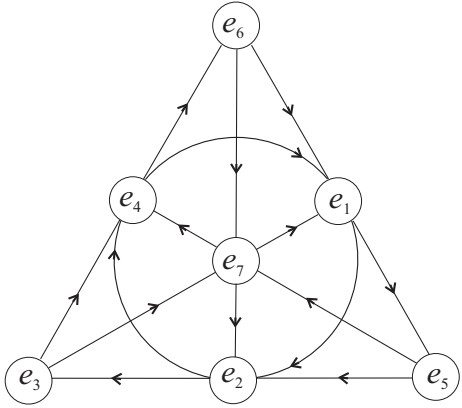, scale=0.4}}
    \caption{\textbf{Fano plane;} the arrows indicate the directions
    of forward cycling; there are unseen arrows closing the cycles,
    for instance between $e_4$ and $e_5$ or $e_3$ and $e_5$.}
    \label{fig:fano}
\end{figure}

The extra imaginary unit $\mat{u}^0$ needed to build a
4-dimensional complex number is built with recourse to matrices
$\mat{E}^{2j} = \rho^2 \lambda^j$. Any combination $\mathrm{i} q_j
\rho^2 \lambda^j/n_q$, with $q_j$ equal to $0$ or $1$ and
$(n_q)^2$ the number of non-zero $q_j$ produces a suitable
imaginary unit. Such combinations bear strong links with octonions
\cite{Baez02}: Assuming justifiable special rules on matrix
multiplication that will ensure non-associativity as required by
the algebra, we can define an octonion basis as follows
\begin{equation}
\begin{split}
    \label{eq:octonbasis}
    e_1 &= \mathrm{i} \rho^2 \lambda^1, \\
    e_2 &= \mathrm{i} \rho^2 \lambda^2, \\
    e_3 &= \mathrm{i} \rho^2 \lambda^3, \\
    e_4 &= \frac{\mathrm{i}\rho^2 \left(\lambda^1 + \lambda^2 \right)}
    {\sqrt{2}}, \\
    e_5 &= \frac{\mathrm{i}\rho^2 \left(\lambda^2 + \lambda^3 \right)}
    {\sqrt{2}}, \\
    e_6 &= \frac{\mathrm{i}\rho^2 \left(\lambda^1 + \lambda^2 + \lambda^3
    \right)}{\sqrt{3}}, \\
    e_7 &= \frac{\mathrm{i}\rho^2 \left(\lambda^1 + \lambda^3 \right)}
    {\sqrt{2}}.
\end{split}
\end{equation}
The multiplication table for octonions can be inferred from Fig.\
\ref{fig:fano} where the arrows indicate directions of forward
cycling, i.e. directions where sign is maintained; for instance
$e_1 e_2 = e_4$ or $e_4 e_6 = e_3$. In the reverse direction the
sign of the result is negative.

Using the multiplication table and the basis definition of Eq.\
(\ref{eq:octonbasis}) we can establish the non-associative
multiplication rule for octonions based on the $\alpha$ matrices:
\begin{enumerate}
  \item remove all normalization factors,
  \item perform the matrix addition of the two elements,
  \item replace any occurrence of $2 \lambda^j$ by zero,
  \item renormalize to unit determinant according
  to the number of matrices in the sum,
  \item change sign if multiplying in the reverse direction
  as defined by the Fano plane,
  \item any element squared follows matrix multiplication rules
  and equals $-\mat{I}$.
\end{enumerate}
The justification for this rules is solely found on the fact that
the elements of Eqs.\ (\ref{eq:octonbasis}) equipped with such
product produce an octonion basis. As an example of use of the
rules, we can check that $e_6 e_4 = -e_3$:
\begin{enumerate}
    \item remove normalization factors,

       $e_4~\rightarrow~\mathrm{i}\rho^2(\lambda^1+\lambda^2)$,\\
        $e_6~\rightarrow~\mathrm{i}\rho^2(\lambda^1+\lambda^2
        +\alpha^3)$,
    \item perform addition,

        $\mathrm{i}\rho^2(2 \lambda^1 + 2 \lambda^2 + \lambda^3)$,
    \item replace factors of $2$ by zero,

        $\mathrm{i}\rho^2\lambda^3~\rightarrow~e_3$,
    \item no renormalization needed,
    \item change sign for reverse direction $\rightarrow -e_3$,
    \item not applicable.
\end{enumerate}

Under the assumptions above a 4D complex number has the form
\begin{equation}
    \label{eq:u0octonion}
    \psi = \frac{p_0 q_j \rho^2\lambda^j}
    {n_q}\, \exp\left(\frac{\mathrm{i}p_0 q_j \rho^2\lambda^j x^0}
    {n_q}\, +  \mathrm{i}p_k \alpha^k x^k\right).
\end{equation}
These numbers cannot be associated to transformations of 4-space
because they involve all 15 dimensions. Presumably the 15
coordinates are not independent and the actual space dimension is
lower than 15 but this is a subject which needs profound research
which is beyond the scope of the present work. Some insight into
the sort of transformations induced by the $\mathrm{i}\alpha^j$
imaginary units can be gained by considering the following matrix
\begin{equation}
    \label{eq:coordext}
    \left(\begin{array}{cccc}
      x^0+\mathrm{i}x^3 & -x^2 -\mathrm{i}x^1 & 0 & 0 \\
      x^2 -\mathrm{i}x^1 & x^0-\mathrm{i} x^3 & 0 & 0 \\
      0 & 0 & -x^4-\mathrm{i}x^7 & x^6 + \mathrm{i}x^5 \\
      0 & 0 & -x^6 + \mathrm{i}x^5 & -x^0+\mathrm{i}x^7 \
    \end{array}\right).
\end{equation}
Considering all the particular situation $\psi = \exp(\mathrm{i}
\theta \alpha^1/2)$ and applying a similarity transformation
coordinates $x^0$, $x^1$, $x^4$ and $x^5$ are preserved while the
other coordinates are transformed as
\begin{equation}
\begin{split}
    \label{eq:coordtrans}
    {x'}^2 &= x^2 \cos \theta + x^3 \sin \theta, \\
    {x'}^3 &= -x^2 \sin \theta + x^3 \cos \theta, \\
    {x'}^6 &= x^6 \cos \theta - x^7 \sin \theta, \\
    {x'}^7 &= x^6 \sin \theta + x^7 \cos \theta.
\end{split}
\end{equation}
If the other $\mathrm{i}\alpha^j$ units were tested we would find
the sets $(x^1, x^2, x^3)$ and $(x^5, x^6, x^7)$ to undergo
rotations of opposite sign. We are thus led to assume that
coordinates $x^0$ to $x^3$ and $x^4$ to $x^7$ are interrelated,
reducing the space dimension to 11; this is not an unexpected
dimension and is coincident with what is used by superstring
theory \cite{Greene99}. Transformations involving any of the
$\mat{u}^0$ units are more complex, involving all the elements of
the coordinate matrix.

Considering $p_j = 0$ in Eq.\ (\ref{eq:u0octonion}), the unitary
hypercomplex numbers represented through the imaginary units of
Eq.\ (\ref{eq:octonbasis}) have the form
\begin{equation}
    \label{eq:x0spinor}
    \frac{q_j \rho^2\lambda^j
    }{n_q}\,\exp\left(\frac{\mathrm{i}\theta q_j \rho^2\lambda^j
    }{2 n_q} \right),
\end{equation}
where $\theta = 2 p_0 x^0$ is an arbitrary real number. If the
upper left quarter of the coordinate matrix represents 4-space, as
in (\ref{eq:coordext}), these numbers represent 2 coordinate
preserving transformations in the cases of $e_1$, $e_2$, $e_3$, 1
coordinate preserving transformations in the cases of $e_4$,
$e_5$, $e_7$ and no coordinate preserving transformations in the
case of $e_6$. These numbers can be classified according to the
signs of $p$ and $q_j$, and according to the number of non-zero
$q_j$. For each sign of $p$ there is one hypercomplex number for
$(n_q)^2$ equal to 3 and 3 hypercomplex numbers for $(n_q)^2$
either 1 or 2.

There is another unexplored possibility for the 0th imaginary unit
which consist of making $\mat{u}^0 = \mathrm{i}\alpha^0$; this
generates yet a different sort of transformation analogous to a
global harmonic pulsation of the whole 4-space. Consider $\psi =
\exp(\mathrm{i}\alpha^0 \theta/2$ and apply a similarity
transformation to the coordinate matrix; all the elements in the
upper left quarter, those representing 4-space, will appear
multiplied by $\cos \theta$.

These generalized complex numbers will be used in the present and
following sections to write solutions of the first order and
source equations. We may have used some discretion in choosing the
mathematical entities we defined as 4D complex numbers; we were
guided in this choice by need to make a connection with elementary
particles later on.

The source equation (\ref{eq:founding}) can become extremely
intricate; this is to be expected, since it is applicable to all
the problems of dynamics. The exposition of this section will be
made clearer by a preliminary exploration of the 4D Helmholtz
equation's solutions using the 4D complexes introduced above.
Consider Eq.\ (\ref{eq:3dwave}) with $j$ and $k$ indices replaced
by $\mu$ and $\nu$ in order to make it 4-dimensional
\begin{equation}
    \label{eq:4dwave}
    \delta^{\mu \nu} \partial_{\mu \nu} \psi = -m^2 \psi.
\end{equation}
Just as in the 3D case, this equation accepts complex solutions
of the type $\psi = |\psi| \exp(\pm \mathrm{i} p_\mu x^\mu)$, with
the condition
\begin{equation}
    \label{eq:4dcond}
    \delta^{\mu \nu}p_\mu p_\nu = m^2.
\end{equation}
We are interested in exploring 4D complex solutions of the form
given by Eq.\ (\ref{eq:4dcomplex}); no matter which choice is made
for $\mathbf{\mathrm{u}}^0$, it will always be possible to write a
solution in the form of Eq.\ (\ref{eq:nonstatpart}) with Condition
(\ref{eq:4dcond}) verified. This solution represents a spin 1/2
wave in the direction of the unit 4-vector $\hat{\mathrm{p}}$
whose components are $p_\mu / m$; $m$ is also the wave frequency
along the direction $\hat{\mathrm{p}}$; in 4D we will associate
the designation \emph{Compton frequency} to the wave frequency.
The 4D wave is made up of two components: One 3-space component
generated by the 3 elements $p_j$, with frequency $\sum (p_j)^2$,
corresponding to the \emph{De Broglie frequency}, and one
component in the direction $x^0$, with frequency $p_0$,
corresponding to a transformation that does not preserve $x^0$.
When $\mat{u}^0 = \mathrm{i}\alpha^0$ the $x^0$ wave induces a
pulsation of 4-space, as referred.

In this section we want to explore the simplest possible forms of
the source equation, together with its solutions. The form of the
equation is governed by the metric which includes an inertial
component and a field component; we will assume here the latter to
be the identity matrix, so we will limit ourselves to identifying
the recursion mechanism which determines the inertia component.

If we consider 4D complexes defined by Eq. (\ref{eq:4dcomplex}), a
subsequent application of Eqs.\ (\ref{eq:recursion}) and
(\ref{eq:smetric}) generates an anti-symmetric metric given by the
following relations
\begin{equation}
\begin{split}
    \label{eq:smunu}
    g^{00} &= \frac{|\psi| |\psi|}{||\psi||^4} =
    \frac{1}{||\psi||^2}, \\
    g^{0j} &= \frac{|\psi| \alpha^j}{||\psi||^2}, \\
    g^{j0} &= \frac{\alpha^j |\psi|}{||\psi||^2}, \\
    g^{jj} &= \frac{1}{||\psi||^2},\\
    g^{jk} &= - g^{kj} = {\alpha^j \alpha^k}{||\psi||^2}
\end{split}
\end{equation}
If $|\psi|$ anti-commutes with $\alpha^j$ this is a situation
where the metric is anti-symmetric and can be replaced by a
diagonal metric; making the substitution $||\psi||^2\rightarrow
m^2$ one gets
\begin{equation}
    \label{eq:freetens}
    g^{\mu \nu} = \frac{1}{m^2}\, \delta^{\mu \nu},
\end{equation}
as the metric to be used in the source equation.

Leaving aside the anti-symmetry conditions, having assumed the
norm to be constant, and noting that a constant metric zeroes the
Christoffel symbols allowing the covariant derivatives to be
replaced by partial ones, the source equation (\ref{eq:founding})
becomes the 4D Helmholtz equation (\ref{eq:4dwave}). Apart from a
phase factor, this equation allows solutions of the type given by
Eq. (\ref{eq:4dcomplex}) with Condition (\ref{eq:4dcond}) met and
with diagonal modulus. If we want not only the source equation but
also the first order equation (\ref{eq:1storder}) to be verified,
then the modulus is given by Eq.\ (\ref{eq:cond0}).

\section{\label{particles}A Physical Evaluation of
the Massive Particle Solutions}
What is specially interesting about 4-dimensional standing waves
in comparison to their 3D counterparts is that there are different
sorts, according to the imaginary unit $\mathbf{\mathrm{u}}^0$
that is chosen. Looking at the octonion basis
(\ref{eq:octonbasis}), elements $(e_1, e_2, e_3)$ identify
transformations where 2 of the 3-space coordinates are preserved,
elements $(e_4, e_5, e_7)$ refer to transformations preserving
only one 3-space coordinate and element $e_6$ 4D combined
transformations involving all 3-space coordinates. For the 7
possible $\mathbf{\mathrm{u}}^0$ choices that form an octonion
basis the concept of octonion units can be extended to the waves
they generate, suggesting that octonion products performed among
4D waves may describe particles involved in chromodynamics. The
different possible choices for $\mathbf{\mathrm{u}}^0$, including
$\mathrm{i}\alpha^0$, point to the interpretation that Eq.\
(\ref{eq:nonstatpart}) represents 8 particles and their respective
anti-particles, with this solution describing the known elementary
particle dynamics. While the equations have so far been written as
pure numbers, dimensions and physical interpretation are possible
through the use of \emph{dimensional factors} introduced in Ref.\
\cite{Almeida02:3}.

Returning to Eq.\ (\ref{eq:nonstatpart}), assume that $p_1 = p_2
=0$ so that the particle moves along the $x^3$ direction. We have
$m^2 = (p_0)^2 + (p_3)^2$ and the first thing we notice is that
the frequency $(p_0)$ of the $x^0$ wave is reduced as $(p_3)$
increases, i.\ e.\ the transformation involving $x^0$ with 1, 2 or
3 spatial coordinates, is transferred to rotation around $x^3$.
Secondly we notice that the De Broglie frequency is $p_3$ with
spin frequency of twice that value; we refer to this fact saying
that the particle has spin $1/2$; actually the spin can also be
$-1/2$ when the sign factor $s$ is negative. Rewriting Eq.\
(\ref{eq:nonstatpart}) for a \emph{stationary} particle, defined
by $p_j=0$, whose $\mathbf{\mathrm{u}}^0$ is chosen among the
octonion basis elements
\begin{equation}
    \label{eq:statpart}
    \psi =  \frac{m q_j \rho^2\lambda^j}{n_q}\,
    \exp\left(\frac{s \mathrm{i} m q_j \rho^2\lambda^j}
    {n_q}\right),
\end{equation}
we identify the particle's mass as $m$, connecting it with the
norm of the particle's wave function; for a stationary particle
the mass coincides with the frequency of the $x^0$ wave.

The particle's electric charge can be defined as $\sum q_j/3$ and
comes from the transformation involving the 0th coordinate through
one octonion unit. For this reason, when $(n_q)^2$ equals 1 or 3,
we will choose negative rather than positive $q_j$, thereby
allowing the electric charge to be $-1$ when $n_q^2=3$. Such a
particle can be then associated with an electron.

The electric charge is $+2/3$ when $n_q^2=2$ and $-1/3$ when
$n_q^2 = 1$, corresponding to the up and down quarks, respectively
\cite{Barnett00, Cottingham98}. Color charge is associated with
the various $q_j \neq 0$ possibilities and we can see that
electric charge appears as a spherically symmetric color charge.
Anti-particles are in turn represented by an equation identical to
Eq.\ (\ref{eq:statpart}) where the sign of the $q_j$'s is
reversed. The interpretation made above shows that Eq.\
(\ref{eq:statpart}) has all the necessary features to represent
all leptons and quarks with the exception of neutrinos, with the
particle's mass being the only free parameter. It is of course
desirable to remove all free parameters, deriving mass from some
characteristics of the source equation's solutions, but we will
not address the subject in this paper.

We are going to associate neutrinos with the
$\mathbf{\mathrm{u}}^0= \mathrm{i} \alpha^0$ situation; as for the
other particles, the neutrinos' masses appear as free parameters.
\section{Solutions with zero norm -
The Question of Massless Particles \label{0norm}}
It is important to investigate solutions of Eq.\ (\ref{eq:4dwave})
when $m=0$ noting that the equation can be rearranged as
\begin{equation}
    \label{eq:0mass1}
    \delta^{jk} \nabla_{jk} \psi = -\nabla_{00} \psi.
\end{equation}
Whenever
\begin{equation}
    \label{eq:d00psi}
    \nabla_{00}\psi = \omega^2 \psi
\end{equation}
and covariant derivatives can be replaced by partial ones, the
equation above becomes the 3D Helmholtz equation
(\ref{eq:3dwave}), for which there are plane wave solutions of the
type given by Eq.\ (\ref{eq:3dsol1}), i.\ e.\ no-spin waves. Being
in 4D, integration of Eq.\ (\ref{eq:d00psi}) has to be made with
$\psi$ as a $4 \times 4$ matrix; after integration we get
\begin{equation}
    \label{eq:0integrate}
    \psi = \psi_{1,2,3} \mathrm{e}^{\pm \mat{h} \omega x^0},
\end{equation}
where $\mat{h}$ is an Hermitian unit and $\psi_{1,2,3}$ is a $4
\times 4$ matrix function of the 3 spatial coordinates. The
general form of a plane wave solution for Eq.\ (\ref{eq:0mass1})
can be written
\begin{equation}
    \label{eq:0masssol}
    \psi = \mat{C} \mathrm{e}^{\pm \mat{h}\omega x^0}
    \mathrm{e}^{\mathrm{i} s p_j x^j};
\end{equation}
with $\mat{C}$ an Hermitian matrix integration constant.

Physical interpretation of Eq.\ (\ref{eq:0masssol}) reveals a
plane wave restricted to 3D space and evanescent in the 0th
dimension, if the positive sign is rejected. The phenomenon of
evanescent waves is well known from total reflection and
waveguides, where both exponent signs are also present but the
positive sign is ignored for lack of physical significance. Here
we interpret the solutions given by Eq.\ (\ref{eq:0masssol}) as
representing massless particles, of which we distinguish three
different types according to the Hermitian unit $\mat{h}$ that is
present: An Hermitian unit obtained from the octonion basis
element $e_6$ by product with $\mathrm{i}$ carries electric charge
in the evanescent wave and will be associated to photons; an
Hermitian unit obtained from one of the other octonion basis
elements by product with $\mathrm{i}$ carries color charge in the
evanescent wave and will be associated to gluons; finally the
Hermitian unit $\alpha^0$ carries no charge and will be discussed
later in this section.

Some words are needed here about the reason why we chose no-spin
wave solutions for Eq.\ (\ref{eq:0mass1}) while spin 1/2 solutions
exist. We could argue that spin 1/2 solutions cannot coexist in
the ground state due to the Pauli exclusion principle, but we can
do better than that: A no-spin wave defines a direction in 4-space
but not a worldline; this is true both for dynamic or observer
space. On the contrary, the observer space appearance of a spin
1/2 wave is similar to a 4-dimensional waveguide and it defines a
worldline coinciding with the waveguide axis; this idea was first
suggested in Ref.\ \cite{Almeida02:1}. The number of cycles along
any waveguide length is actually a measure of time interval in the
waveguide's (or particle's) own clock. We can also assume all
clocks in all waveguides to be synchronized from a common "zero
time" in the origin of the Universe. When two waveguides meet in
the same 3D position and their time measurement coincides they
have to interact; Ref.\ \cite{Almeida02:2} includes the discussion
of a collision problem in 4-dimensional optics, which is an
illustration of this fact. The same does not happen with no-spin
waves which never actually meet in one position because they are
not associated to lines and this matches the behavior of massless
particles. The interpretation of time as the count of cycles along
the waveguide is entirely compatible with the use of $\mathrm{d}t$
as line element that we make in the next section.

A superposition of an even number of $\pm 1/2$ spin waves has many
characteristics of a no-spin wave; consider for instance the
superposition of 2 equal modulus equal frequency $+1/2$ spin waves
with different phase, both spinning around the $x^1$ direction:
\begin{equation}
    \label{eq:2waves}
    \psi = |\psi|\left[ \mathrm{e}^{\mathrm{i}\alpha^1 \omega x^1}
    + \mathrm{e}^{\mathrm{i}\alpha^1 (\omega x^1 + \rho)}\right];
\end{equation}
it is sufficient to make $\rho = \pi$ for the equation above to
degenerate into a no-spin wave given by $\psi = |\psi| \cos(\omega
x^1)\mat{I}$; naturally the same could be done if the two waves
were counter-spinning or with any even number of waves co- or
counter-spinning. The spin of any wave will thus be expressed by
the algebraic sum of the superposed waves spins.

All solutions verifying Eq.\ (\ref{eq:0masssol}) have complex
dependence on the spatial coordinates, not quaternionic-like
solutions. As we said before, complex numbers describe no-spin
waves, whereby these solutions all exhibit no spin. Only
combinations of an even number of \emph{particles} with spin 1/2
can then lead to integral spin waves, behaving like no-spin ones.
We shall examine this case next.

We know from experiment that photons are associated to the
annihilation of an electron and a positron; since we have
established the wavefunctions of the intervening particles it is
elucidating to investigate how the annihilation process can be
understood in terms of those wavefunctions. The wavefunctions of
an electron and a positron are given by
\begin{equation}
\begin{split}
    \label{eq:electpos}
    \psi_- &= \mathrm{i}m e_6 \mathrm{e}^{-s m e_6 x^0
    + \mathrm{i} p_j\alpha^j x^j},\\
    \psi_+ &= \mathrm{i}m e_6 \mathrm{e}^{s m e_6 x^0
    + \mathrm{i} p_k\alpha^k x^k};
\end{split}
\end{equation}
where $e_6$ is the octonion basis element corresponding to
electron and positron and the exponent sign designates each
particle's electric charge; the sign of $s$ stands for the two
possible spin orientations. By adding an $\mathrm{i} m e_6 x^0$
term within both exponents to represent the coupling between spins
and multiplying the two wave functions we get
\begin{align}
    \label{eq:annihilate1}
    \psi_- \otimes \psi_+ &\Rightarrow m^2
    \mathrm{e}^{-s m e_6 x^0 (1-\mathrm{i})}
    \otimes  \mathrm{e}^{s m e_6 x^0 (1+\mathrm{i})} \\
    &\Rightarrow m^2 \mathrm{e}^{2 s m \mathrm{i}e_6 x^0}
    \left(2 * \mathrm{e}^{\mathrm{i} p_j\alpha^j x^j}\right)
    \tag{\ref{eq:annihilate1}$'$} \\
    &\Rightarrow m \mathrm{e}^{ s m \mathrm{i}e_6 x^0}
    \left(2 * \mathrm{e}^{\mathrm{i} p_j\alpha^j x^j}\right)\notag\\
    &\quad \otimes m \mathrm{e}^{ s m \mathrm{i}e_6 x^0}
    \left(2 * \mathrm{e}^{\mathrm{i} p_j\alpha^j x^j}\right).
    \tag{\ref{eq:annihilate1}$''$}
\end{align}
In the first stage of the process (\ref{eq:annihilate1}), the two
particles are assumed to interact by means of the $\mathrm{i}x^0$
terms in the exponents; in the second stage
(\ref{eq:annihilate1}$'$) an intermediate particle is formed,
which absorbs the result from the annihilation in the $p_j$; the
superposition of the two waves is indicated by $2 *$. Finally the
third stage (\ref{eq:annihilate1}$''$) produces 2 identical
photons of spin $1$ and frequency equal to the mass of one
particle. The final result is a solution of Eq.\
(\ref{eq:0mass1}), representing a superposition of 3D spin waves,
evanescent in the $x^0$ direction. As before only a negative
evanescent exponent is acceptable and since $s$ can have both
signs the sense of the mutual orbiting must change according to
the spin sign.

The annihilation of two neutrinos, if at all possible, leads to a
similar expression, where $\mathrm{i}e_6$ is replaced by
$\alpha^0$; the resulting particles are
\begin{equation}
    \label{eq:neutrinoanni}
    \psi = m \mathrm{e}^{- m \alpha^0 x^0}
    \left(2 * \mathrm{e}^{\mathrm{i} p_j\alpha^j x^j}\right).
\end{equation}
This would be a massless particle not involved in any of the four
known fundamental interactions and we do not dare proposing an
interpretation for it, although it might be related to dark matter
in the Universe.

Returning to Eq.\ (\ref{eq:annihilate1}), further annihilation is
possible through the interaction of two $\psi_- \psi_+$ pairs with
opposing evanescent exponent signs $s$. We rejected the positive
sign for its lack of physical significance, however two pairs with
opposite signs produce pure 3-dimensional waves, not accompanied
by any $x^0$ evanescence; these waves are the superposition of 4
wavefunctions and thus have spin $2$. Four identical particles
with mass $m$ can be annihilated producing 4 resulting particles
represented by
\begin{equation}
    \label{eq:graviton}
    \psi = m \left(4 * \mathrm{e}^{\mathrm{i} p_j\alpha^j x^j}\right).
\end{equation}

We believe these are good candidates for gravitons which may have
been detected experimentally long ago by Allais' measurements of
Foucault's pendulum anomalies \cite{Allais59} and already
discussed by the author \cite{Almeida01:5}.
\section{Solutions with fields}
In sections \ref{particles} and \ref{0norm} we discussed field
free solutions of the source equation in situations which reduced
that equation to the 4D Helmholtz equation; those solutions where
found to have characteristics which suggested their physical
interpretation as massive and massless particles. However the real
test of this interpretation relies on the ability of the particles
thus defined to react properly to applied fields, accurately
predicting the dynamics of those particles. Fields are assumed to
alter the particle's dynamic space through the $\mat{s}^\mu$
matrices defined in Eq.\ (\ref{eq:recursion}). The exact way in
which fields are created is not yet clearly understood but we
believe it is connected to the flux of massless particles; we
think that the flux of massless particles determines the field
gradient, the net value of the flux being linked to the norm of
the field gradient and the flux direction to its direction. Things
get more complicated when massless particles are originated in
moving frames because an extra rotational factor intervenes. The
whole process will eventually be governed by equations similar to
Maxwell's equations albeit generalized to encompass the three
field types: gravitational, electromagnetic and color. In the
following exposition fields are presented axiomatically and the
dynamics they induce are then seen to represent dynamics under the
influence of gravity and electromagnetism; a loose association is
made with the massless particles identified in the previous
section but no attempt is made to propose a formal link between
particle flux and field gradient, which we intend to do in a
future paper.

We are going to assume that fields leave the particle wavefunction
$\psi$ virtually unaltered, so that we can use the wavefunctions
derived previously; this means that we can make $|\psi|=
\mathrm{i}m\mat{u}^0$ and $||\psi||=m$. It will frequently be
convenient to write the modulus in the equivalent form $|\psi| =
\mathrm{i}m \exp(\pm \mat{u}^0 \pi/2)$; as a first generalization
step the $\mat{s}^\mu$ matrices are modified from Eq.\
(\ref{eq:recursion}) by introducing $G_\mu$, scalars, and
$\mat{A}_\mu$, matrices:
\begin{equation}
    \label{eq:recursfield}
    \mat{s}^0 = \frac{\mathrm{i}G_\mu \alpha^\mu \mathrm{e}^{\pm\mat{u}^0
    \mat{A}_\mu \pi/2}}{m},
    ~~~~\mat{s}^j =
    \frac{G_\mu \alpha^j}{m}.
\end{equation}
Eq.\ (\ref{eq:1storder}) maintains its validity and by applying to
both members the operator $\mathrm{i}\mat{s}^\mu \nabla_\mu$ we
get the source equation (\ref{eq:founding}) with $g^{\mu \nu}$
defined by Eq.\ (\ref{eq:smetric}) as before.

In a general situation the metric defined by Eq.\
(\ref{eq:smetric}) can be expanded to
\begin{equation}
\begin{split}
    \label{eq:fieldmetric}
    g^{00} &= \frac{-1}{m^2}\displaystyle{ \sum_\mu (G_\mu)^2
    \mathrm{e}^{\pm \mat{u}^0
    \mat{A}_\mu \pi}}, \\
    g^{0j} &= \frac{\mathrm{i}}{m^2}G_\mu\alpha^\mu\mathrm{e}^{\pm\mat{u}^0
     \mat{A}_\mu \pi/2} G_j \alpha^j, \\
    g^{j0} &= \frac{\mathrm{i}}{m^2}G_j  \alpha^j G_\mu \alpha^\mu\mathrm{e}^{\pm
    \mat{u}^0\mat{A}_\mu \pi/2},  \\
    g^{jk} &= \frac{1}{m^2}G_j \alpha^j G_k \alpha^k.
\end{split}
\end{equation}
This metric will now have to be examined for various possible
forms of $G_\mu$ and $\mat{A}_\mu$ and the respective dynamics
related to known interactions. Simultaneously a connection will be
made to the massless particles identified in the previous section
in order to associate them with mediators of the various
interactions.

The simplest case to examine is of course the one where all the
$\mat{A}_\mu$ are zero and the $G_\mu$ coefficients are all equal,
$G_\mu = G/4$; inserting into Eq.\ (\ref{eq:fieldmetric}), the
metric becomes anti-symmetric and we can reduce it to the diagonal
metric
\begin{equation}
    \label{eq:gravmetr}
    g^{\mu \nu} = \left(\frac{G}{m}\right)^2 \delta^{\mu \nu}.
\end{equation}
This metric defines a space whose geodesics reproduce the
dynamics of a particle with mass $m$ in a gravitational field
$G$, as long as the field sources are stationary. This problem
was analyzed in Ref.\ \cite{Almeida02:3} and will be reproduced
below for completeness.

Inserting Eq.\ (\ref{eq:gravmetr}) into Eq.\ (\ref{eq:founding}),
with $\psi$ given by Eq.\ (\ref{eq:4dcomplex}) and assuming $G$ to
be a slowly varying function so that the Christoffel symbols can
be neglected
\begin{equation}
    \label{eq:gravity1}
    \left(\frac{G}{m}\right)^2 \delta^{\mu \nu} p_\mu p_\nu =1;
\end{equation}
then replacing $p_\mu = g_{\mu \nu} \dot{x}^\nu$, with
$\dot{x}^\mu = \mathrm{d}x^\mu/\mathrm{d}t$ the derivative of the
coordinate with respect to the geodesic line element, here
designated by $\mathrm{d}t$,
\begin{equation}
    \label{eq:gravity2}
    \left(\frac{m}{G} \right)^2 \delta_{\mu \nu}
    \dot{x}^\mu \dot{x}^\nu = 1.
\end{equation}
Here we remind the reader that although we have chosen the letter
$t$ for the line element, this is in no way associated with time
at this stage; the procedure whereby this association can be made
later on was detailed in Ref.\ \cite{Almeida02:3}.

The equation above is called a dynamic space equation because the
inertia element $m$ is present in the metric; it can be converted
into an observer space equation by the coordinate change $X^\mu =
m x^\mu$. We can then write
\begin{equation}
    \label{eq:gravity3}
    \frac{\delta_{\mu \nu}
    \dot{X}^\mu \dot{X}^\nu}{G^2} =1.
\end{equation}

The procedure to derive the 4 geodesic equations from Eq.\
(\ref{eq:gravity3}) can be found in many textbooks, for instance
\cite{Martin88}. We define a constant Lagrangian with the value
$1/2$ and insert into the second member of Eq.\
(\ref{eq:gravity3}) which becomes $2L$. Following that, we derive
the Euler-Lagrange equations in a standard way. If $G$ is
independent from coordinate $X^0$ there is conservation of the
corresponding conjugate momentum
\begin{equation}
    \label{eq:p0conserv}
    \frac{\dot{X}^0}{G^2} = \frac{1}{\gamma},
\end{equation}
with $\gamma$ an arbitrary constant. For the other 3 coordinates
we can write
\begin{equation}
\begin{split}
    \label{eq:pjderiv}
    \frac{\mathrm{d}}{\mathrm{d}t} \left(\frac{\dot{X}^j}
    {G^2}\right) &= -\frac{\delta_{\mu \nu}
    \dot{X}^\mu \dot{X}^\nu \partial_j G}
    {G^3}\\
    &= -\frac{\partial_j G}{G},\qquad
    \text{using (\ref{eq:gravity3})}.
\end{split}
\end{equation}
The reader is now referred to Ref.\ \cite{Almeida02:3} in order to
verify that in the particular case of a single body of mass $M$ as
origin of the  field introduced in Eq.\ (\ref{eq:recursfield}),
one must make $G = \exp(-M/r)$, with $r$ the distance between the
two masses, thereby giving a clue that $\ln G$ is the Newtonian
gravitational field. In the cited reference it was also
demonstrated that predictions of general relativity are verified
to the first order approximation by the equations above.

The link of gravitational interaction introduced in Eq.\
(\ref{eq:recursfield}) with the graviton equation
(\ref{eq:graviton}) can be made by noting that both $G$ and
graviton's modulus are products of the identity matrix by a
scalar; it is thus arguable that the gravitational field can be
mediated by such particles.

Next we want to investigate the adequacy of function $\psi$ to
represent the electrodynamics of a charged particle. We start by
taking $G$ as unity and
\begin{equation}
    \label{eq:emfield}
    \mat{A}_\mu = e_6 A_\mu,
\end{equation}
with $A_\mu$ scalar; we will also assume that $\mat{u}^0 = \pm
e_6$, effectively limiting the analysis to the electron/positron
family members. As a result the exponentials $\exp(\pm\mat{u}^0
\mat{A}_\mu \pi/2)$ in Eq.\ (\ref{eq:fieldmetric}) can be replaced
by $\exp(\pm A_\mu \pi/2)$. It is also convenient to define
\begin{equation}
    \label{eq:vpot}
    V_\mu = \mathrm{e}^{\pm A_\mu \pi/2}.
\end{equation}
The metric elements become
\begin{equation}
\begin{split}
    \label{eq:emmetric}
    g^{00} &= \frac{\delta^{\mu \nu}V_\mu V_\nu}{m^2}, \\
    g^{0j} &= \frac{V_\mu \alpha^\mu \alpha^j}{m^2}\\
    g^{j0} &= \frac{V_\mu \alpha^j \alpha^\mu}{m^2}\\
    g^{jk} &= \frac{\alpha^j \alpha^k}{m^2}.
\end{split}
\end{equation}
This is no longer an anti-symmetric metric such as we have found
in all previous cases because $g^{0j}$ and $g^{j0}$ do not cancel
each other completely. After cancellation of the anti-symmetric
parts the metric becomes
\begin{equation}
    \label{emmetric2}
    g^{\mu \nu} = \frac{1}{m^2}\left(\begin{array}{cccc}
      \delta^{\kappa \lambda} V_\kappa V_\lambda
      &  V_1 & V_2
        & V_3 \\
      V_1 & 1 & \cdot & \cdot \\
      V_2 & \cdot & 1 & \cdot \\
      V_3 & \cdot & \cdot & 1 \
    \end{array} \right).
\end{equation}
The lower subscript metric $g_{\mu \nu}$ is obtained, as usual,
by calculating the inverse of $g^{\mu \nu}$
\scriptsize
\begin{equation*}
    \label{eq:eminverse}
    \frac{m^2}{(V_0)^2}\left(\begin{array}{cccc}
      1 & -V_1
      & -V_2 & -V_3 \\
      -V_1 & (V_0)^2 +(V_1)^2
      & V_1 V_2 & V_1 V_3 \\
      -V_2 & V_1 V_2
      & (V_0)^2 + (V_2)^2 & V_2 V_3 \\
      -V_3 & V_1 V_3
      & V_2 V_3 & (V_0)^2 + (V_3)^2 \
    \end{array}\right).
\end{equation*}
\normalsize We don't develop here the details of electrodynamics
generated by the metric above; the procedure is similar to what
was used for gravity and can be found in Ref.\
\cite{Almeida02:3}, where both electric and magnetic field
dynamics were examined.

Consider now the situation where $\partial_{\mu} V_\nu = 0$ except
for $\partial_{1} V_2 = q B_3$; for simplicity we make $V_0 =1$,
$V_1 = V_3 = 0$ and $V_2 = x^{1} q B_3$; $B_3$ represents the
magnetic field aligned along $x^3$ and $q=\pm 1$. The Lagrangian
is now defined by
\begin{equation}
    \label{eq:maglag}
    \frac{2L}{m^2} = \delta_{\mu\nu}\dot{x}^\mu \dot{x}^\nu
     -2q x^1 B_3 \dot{x}^0 \dot{x}^2
    + \left(q x^1 B_3 \dot{x}^2 \right)^2.
\end{equation}

Some straightforward calculations lead to the conjugate momenta
\begin{align}
    \label{eq:magmom0}
    & \frac{p_0}{m^2} = \dot{x}^0
    -q x^1 B_3 \dot{x}^2,\\
    \label{eq:magmom1}
    & \frac{p_1}{m^2}= \dot{x}^1,\\
    \label{eq:magmom2}
    & \frac{p_2}{m^2}= \dot{x}^2
    -q x^1 B_3 \dot{x}^0
    + \left(q x^1 B_3 \right)^2
    \dot{x}^2,\\
    \label{eq:magmom3}
    & \frac{p_3}{m^2}= \dot{x}^3.
\end{align}
Since the Lagrangian is independent from $x^3$, Eq.\
(\ref{eq:magmom3}) gives
\begin{equation}
    \label{eq:mag3}
    \dot{x}^3 = \text{constant};
\end{equation}
then, independence of the Lagrangian from $x^0$ and Eq.\
(\ref{eq:magmom0}) give
\begin{equation}
    \label{eq:mag0}
    \dot{x}^0 = \frac{1}{m \gamma} +
    q x^1 B_ 3 \dot{x}^2,
\end{equation}
with $\gamma$ a constant greater than $1$, unity applying to a
stationary particle.

Inserting Eq.\ (\ref{eq:mag0}) into Eq.\ (\ref{eq:magmom2}) and
deriving the second member with respect to the line element
$\mathrm{d}t$ and equating to zero because the Lagrangian is
independent from $x^2$
\begin{equation}
    \label{eq:mag2}
    \ddot{x}^2 =
    \frac{q B_3}{m\gamma}\, \dot{x}^1.
\end{equation}
Finally, from Eq.\ (\ref{eq:magmom1}), deriving the Lagrangian
with respect to $x^1$ and making the needed substitutions we get
\begin{equation}
    \label{eq:mag1}
    \ddot{x}^1 =-
    \frac{q B_3}{m\gamma}\, \dot{x}^2.
\end{equation}
Equations (\ref{eq:mag2}) and (\ref{eq:mag1}) represent the
Lorentz force exerted by the magnetic field $B_3$ over a particle
of electric charge $q$ and moving with velocity $\dot{x}^j$. The
frequency $q B_3/( m \gamma)$ is the angular frequency of the
movement.

In the limit $\gamma=1$ and $\dot{x}^j=0$ the particle approach
breaks down because this approach is similar to the geometric or
ray approach of optics and cannot be used when any features of the
worldline are of the same order of magnitude as the particle's
Compton wavelength. If we take $q = -1$, for the electron case,
the angular frequency becomes $\omega = -B_3/(m \gamma)$ and one
possible solution for Eqs.\ (\ref{eq:mag2}) and (\ref{eq:mag1}) is
\begin{equation}
\begin{split}
    \label{eq:electmag}
    x^1 &= r \cos \left(\omega t \right), \\
    x^2 &= r \sin \left(\omega t \right).
\end{split}
\end{equation}
The use of $t$ as an independent variable is here linked to a
specific worldline in dynamic space; $t$ can always be expressed
in terms of the coordinates by inversion of the equations above $t
= \arctan(x^2/x^1)/\omega$. Inserting Eqs.\ (\ref{eq:electmag})
into Eqs.\ (\ref{eq:magmom0}) to (\ref{eq:magmom3}) we get the
conjugate momenta
\begin{equation}
\begin{split}
    \label{eq:conjmomenta}
    p_0 &= \frac{m}{\gamma}, \\
    p_1 &= \frac{m B_3 r \sin \left(\omega t \right)}{\gamma}, \\
    p_2 &= 0, \\
    p_3 &= \text{arbitrary constant.}
\end{split}
\end{equation}
Choosing zero for $p_3$ the particle's wavefunction can be written
\begin{equation}
\begin{split}
    \label{eq:magwave}
    \psi &= \exp \left(\frac{m}{\gamma} \mat{u}^0 x^0
    + \frac{m B_3 r \sin \left(\omega t \right)}{\gamma}\,
    \mathrm{i}\alpha^1 x^1 \right) \\
    &= \exp \left[\frac{m}{\gamma} \mat{u}^0 x^0 +  \mathrm{i}
    \alpha^1 m^2
    \omega r^2 \sin^2 (\omega t) \right];
\end{split}
\end{equation}
which is now an expression not involving $\dot{x}^j$ and not
breaking down when $\gamma \rightarrow 1$. In the limit there is
spin precession with the cyclotron frequency $\omega = -B_3/m$;
for a positron the frequency would show a positive sign.

Similarly to the relation of gravitons with the gravitational
field, photons can be seen as related to the EM field by their
modulus when comparing Eq.\ (\ref{eq:emfield}) with Eq.\
(\ref{eq:annihilate1}). In both cases there is an exponent which
can interact with the exponent in the wavefunction modulus;
photons can be considered as mediators of this field. Naturally
relations like Eq.\ (\ref{eq:emfield}) can be established for the
other octonion basis elements, originating the color fields and
chromodynamics; this is a subject for a forthcoming paper.

We cannot end this analysis without touching upon the situation
where the $G_\mu$ differ from each other; this must surely be
related to gravity, since we have seen how gravitational field is
modelled by equal coefficients. We presume at this point that
different $G_\mu$'s must be associated with mass currents in a
similar way to what relates the $V_\mu$'s to electrical currents.
This would mean that different $G_\mu$ coefficients model
situations of non-stationary gravitational sources. As we said at
the beginning of this section, it is believed that a set of
coupled differential equations links the components of each field
and that these equations are a generalization of Maxwell's
equations; we will investigate this possibility also in a
forthcoming paper.
\section{Conclusion and further work}
Massive and non-massive wave-particles have been found as
solutions of a 4-dimensional source equation. Those solutions were
identified as 4D standing waves and may be related to the
probability waves of quantum mechanics by a switch from the
pseudo-Euclidean space used in this work to the more traditional
hyperbolic space-time. The character of their dynamical symmetry
has been found to follow that of elementary particles, leaving
mass as the only free parameter. Electric and color charge result
directly from rotational symmetry properties in 4D of standing
waves, and massive particles were shown to verify the laws of
dynamics following geodesics of a suitably defined 4-space. The
physical origin of spin has been identified without appeal to
Relativity principles, and found to result merely from Helmholtz
equation requirements in 4D. The modifications to the source
equation needed to express the complete dynamics were loosely tied
to the character of massless particles, this being an area of
further development as detailed below.

Two complementary lines of future work are envisaged, following
paths only suggested in the present work. The first line will try
to complete the set of recursive equations, deriving fields from
particle densities and currents in dynamic and observer space, and
in particular deriving chromodynamics. As a mere working
hypothesis we believe that a composite field can be defined as
\begin{equation}
    \label{eq:compfield}
    \phi_\mu = G_\mu \alpha_\mu
    \mathrm{e}^{\mat{A}_\mu},~~~~\text{(summation\, is\, suppressed)}
\end{equation}
where $G_\mu$ represents the contribution from gravitational
sources both stationary and non-stationary and $\mat{A}_\mu$
represents both electromagnetic and color fields, as discussed in
the previous section. The composite field will then be related to
the \emph{flux} of massless particles by a composite field tensor
defined by
\begin{equation}
    \label{eq:fieldtensor}
    \mat{F}_{\mu \nu} = \partial_\mu \phi_\mu - \partial_\nu \phi_\nu,
\end{equation}
which is expected to verify relations similar to Maxwell's
equations, namely
\begin{gather}
    \label{eq:maxwell1}
    \partial_\lambda \mat{F}_{\mu \nu} + \partial_\nu \mat{F}_{\lambda \mu} +
    \partial_\mu \mat{F}_{\nu \lambda} =0, \\
    \label{eq:maxwell2}
    \partial^\mu \mat{F}_{\mu \nu} = \mat{J}_\nu.
\end{gather}
These are untested hypothesis and must be seen only as a working
program.

The second line of work will be a search for variable norm
solutions of the source equation. Preliminary work has already
shown that allowance for a variable norm and the consequent
consideration of Christoffel symbols can send the source equation
into some modified forms of Emden-Fowler equation
\cite{Polyanin95} which can determine mass, thus eliminating the
only remaining free parameter.
\section{Acknowledgements}
The author wishes to express his deepest thanks to the independent
scholar Roger Y. Gouin (\texttt{email:\ rgouin@mindspring.com})
for the invaluable and countless emails with criticism and
suggestions; this work owes much to him.

  \bibliography{Abrev,aberrations,assistentes}   

\begin{thebibliography}{18}
\expandafter\ifx\csname natexlab\endcsname\relax\def\natexlab#1{#1}\fi
\expandafter\ifx\csname bibnamefont\endcsname\relax
  \def\bibnamefont#1{#1}\fi
\expandafter\ifx\csname bibfnamefont\endcsname\relax
  \def\bibfnamefont#1{#1}\fi
\expandafter\ifx\csname citenamefont\endcsname\relax
  \def\citenamefont#1{#1}\fi
\expandafter\ifx\csname url\endcsname\relax
  \def\url#1{\texttt{#1}}\fi
\expandafter\ifx\csname urlprefix\endcsname\relax\def\urlprefix{URL }\fi
\providecommand{\bibinfo}[2]{#2}
\providecommand{\eprint}[2][]{\url{#2}}

\bibitem[{\citenamefont{Almeida}(2002{\natexlab{a}})}]{Almeida02:3}
\bibinfo{author}{\bibfnamefont{J.~B.} \bibnamefont{Almeida}},
  \emph{\bibinfo{title}{Unification of classical and quantum mechanics}}, 2002,
  \eprint{physics/0211056}.

\bibitem[{\citenamefont{Cottingham and Greenwood}(1998)}]{Cottingham98}
\bibinfo{author}{\bibfnamefont{W.~N.} \bibnamefont{Cottingham}}
  \bibnamefont{and} \bibinfo{author}{\bibfnamefont{D.~A.}
  \bibnamefont{Greenwood}}, \emph{\bibinfo{title}{An Introduction to the
  Standard Model of Particle Physics}} (\bibinfo{publisher}{Cambrige University
  Press}, \bibinfo{address}{Cambridge, U.K.}, \bibinfo{year}{1998}).

\bibitem[{\citenamefont{Born and Wolf}(1997)}]{Born80}
\bibinfo{author}{\bibfnamefont{M.}~\bibnamefont{Born}} \bibnamefont{and}
  \bibinfo{author}{\bibfnamefont{E.}~\bibnamefont{Wolf}},
  \emph{\bibinfo{title}{Principles of Optics}} (\bibinfo{publisher}{Cambridge
  University Press}, \bibinfo{address}{Cambridge, U.K.}, \bibinfo{year}{1997}),
  \bibinfo{edition}{6th} ed.

\bibitem[{\citenamefont{Almeida}(2002{\natexlab{b}})}]{Almeida02:1}
\bibinfo{author}{\bibfnamefont{J.~B.} \bibnamefont{Almeida}},
  \emph{\bibinfo{title}{Prospects for unification under 4-dimensional optics}},
  2002, \eprint{hep-th/0201264}.

\bibitem[{\citenamefont{Dixon}(1994)}]{Dixon94}
\bibinfo{author}{\bibfnamefont{G.~M.} \bibnamefont{Dixon}},
  \emph{\bibinfo{title}{Division Algebras: Octonions Quaternions Complex
  Numbers and the Algebraic Design of Physics}}, vol. \bibinfo{volume}{290} of
  \emph{\bibinfo{series}{Mathematis and its Applications}}
  (\bibinfo{publisher}{Kluwer Academic Publishers},
  \bibinfo{address}{Dordrecht}, \bibinfo{year}{1994}).

\bibitem[{\citenamefont{Altmann}(1986)}]{Altmann86}
\bibinfo{author}{\bibfnamefont{S.~L.} \bibnamefont{Altmann}},
  \emph{\bibinfo{title}{Rotations, Quaternions, and Double Groups}}
  (\bibinfo{publisher}{Clarendon Press}, \bibinfo{address}{Oxford, England},
  \bibinfo{year}{1986}).

\bibitem[{\citenamefont{{van der Waerden}}(1991)}]{Waerden91}
\bibinfo{author}{\bibfnamefont{B.~L.} \bibnamefont{{van der Waerden}}},
  \emph{\bibinfo{title}{Algebra}}, vol.~\bibinfo{volume}{II}
  (\bibinfo{publisher}{Springer-Verlag}, \bibinfo{address}{N. York, USA},
  \bibinfo{year}{1991}).

\bibitem[{\citenamefont{Sternberg}(1995)}]{Sternberg95}
\bibinfo{author}{\bibfnamefont{S.}~\bibnamefont{Sternberg}},
  \emph{\bibinfo{title}{Group Theory and Physics}}
  (\bibinfo{publisher}{Cambridge University Press},
  \bibinfo{address}{Cambridge, U. K.}, \bibinfo{year}{1995}).

\bibitem[{\citenamefont{Wigner}(1959)}]{Wigner59}
\bibinfo{author}{\bibfnamefont{E.~P.} \bibnamefont{Wigner}},
  \emph{\bibinfo{title}{Group Theory and its Application to the Quantum
  Mechanics of Atomic Spectra}} (\bibinfo{publisher}{Academic Press},
  \bibinfo{address}{N. York, USA}, \bibinfo{year}{1959}).

\bibitem[{\citenamefont{Lounesto}(2001)}]{Lounesto01}
\bibinfo{author}{\bibfnamefont{P.}~\bibnamefont{Lounesto}},
  \emph{\bibinfo{title}{Clifford Algebras and Spinors}}, vol.
  \bibinfo{volume}{286} of \emph{\bibinfo{series}{London Mathematical Society
  Lecture Note Series}} (\bibinfo{publisher}{Cambridge University Press},
  \bibinfo{address}{Cambridge, U.K.}, \bibinfo{year}{2001}),
  \bibinfo{edition}{2nd} ed.

\bibitem[{\citenamefont{Baez}(2002)}]{Baez02}
\bibinfo{author}{\bibfnamefont{J.~C.} \bibnamefont{Baez}},
  \emph{\bibinfo{title}{The octonions}}, \bibinfo{journal}{Bull. Am. Math.
  Soc.} \textbf{\bibinfo{volume}{39}}, \bibinfo{pages}{145}, 2002,
  \eprint{math.RA/0105155}.

\bibitem[{\citenamefont{Greene}(1999)}]{Greene99}
\bibinfo{author}{\bibfnamefont{B.}~\bibnamefont{Greene}},
  \emph{\bibinfo{title}{The Elegant Universe: Superstrings, Hidden Dimensions,
  and the Quest for the Ultimate Theory}} (\bibinfo{publisher}{W. W. Norton \&
  Company Inc.}, \bibinfo{address}{N. York}, \bibinfo{year}{1999}).

\bibitem[{\citenamefont{Barnett et~al.}(2000)\citenamefont{Barnett, Muhry, and
  Quinn}}]{Barnett00}
\bibinfo{author}{\bibfnamefont{R.~M.} \bibnamefont{Barnett}},
  \bibinfo{author}{\bibfnamefont{H.}~\bibnamefont{Muhry}}, \bibnamefont{and}
  \bibinfo{author}{\bibfnamefont{H.~R.} \bibnamefont{Quinn}},
  \emph{\bibinfo{title}{The Charm of Strange Quarks: Mysteries and Revolutions
  of Particle Physics}} (\bibinfo{publisher}{Springer-Verlag},
  \bibinfo{address}{N. York, USA}, \bibinfo{year}{2000}).

\bibitem[{\citenamefont{Almeida}(2002{\natexlab{c}})}]{Almeida02:2}
\bibinfo{author}{\bibfnamefont{J.~B.} \bibnamefont{Almeida}},
  \emph{\bibinfo{title}{K-calculus in 4-dimensional optics}}, 2002,
  \eprint{physics/0201002}.

\bibitem[{\citenamefont{Allais}(1959)}]{Allais59}
\bibinfo{author}{\bibfnamefont{M.~F.~C.} \bibnamefont{Allais}},
  \emph{\bibinfo{title}{Should the laws of gravity be reconsidered? {P}art {I}
  -- {A}bnormalities in the motion of a paraconical pendulum on an anisotropic
  support}}, \bibinfo{journal}{Aero/Space Engineering} pp.
  \bibinfo{pages}{46--52}, September 1959.

\bibitem[{\citenamefont{Almeida}(2001)}]{Almeida01:5}
\bibinfo{author}{\bibfnamefont{J.~B.} \bibnamefont{Almeida}},
  \emph{\bibinfo{title}{A theory of mass and gravity in 4-dimensional optics}},
  2001, \eprint{physics/0109027}.

\bibitem[{\citenamefont{Martin}(1988)}]{Martin88}
\bibinfo{author}{\bibfnamefont{J.~L.} \bibnamefont{Martin}},
  \emph{\bibinfo{title}{General Relativity: A Guide to its Consequences for
  Gravity and Cosmology}} (\bibinfo{publisher}{Ellis Horwood Ltd.},
  \bibinfo{address}{U. K.}, \bibinfo{year}{1988}).

\bibitem[{\citenamefont{Polyanin and Zaitsev}(1995)}]{Polyanin95}
\bibinfo{author}{\bibfnamefont{A.~D.} \bibnamefont{Polyanin}} \bibnamefont{and}
  \bibinfo{author}{\bibfnamefont{V.~F.} \bibnamefont{Zaitsev}},
  \emph{\bibinfo{title}{Handbook of Exact Solutions for Ordinary Differential
  Equations}} (\bibinfo{publisher}{CRC Press}, \bibinfo{address}{USA},
  \bibinfo{year}{1995}).

\end{thebibliography}

\end{document}